\documentstyle[12pt,graphics]{article}
\begin{document}

\begin{center}
{\Large\bf Multidimensional Inhomogeneous Cosmology in
Scalar Tensor Theory}\\[20 mm]
D. Panigrahi\footnote{Relativity and Cosmology Research Centre, Jadavpur University,
Kolkata - 700032, India ,E-mail: panigrahi\_dp@yahoo.co.in ,
Permanent Address :
 Kandi Raj College, Kandi, Murshidabad 742137, India},
 Y. Z. Zhang\footnote{Permanent Address :
 Institute of theoretical Physics, Chinese Academy of Sciences,
 P.O.B. 2735, Beijing, China}
  and S. Chatterjee\footnote{Relativity and Cosmology Research Centre, Jadavpur University,
Kolkata - 700032, India, E-mail :
sujit@juphys.ernet.in , Permanent Address :
 New Alipore College, Kolkata  700053, India\\[2mm] Correspondence to: S. Chatterjee
 (sujit@juphys.ernet.in)}\\[10mm]

\end{center}

\begin{abstract}
Exact  cosmological  solutions are obtained for a five dimensional
inhomogeneous  fluid distribution along with a Brans-Dicke type of
scalar   field. The set includes  varied  forms  of  matter  field
including $\rho+p=0$,  where  p     is     the     3D    isotropic
pressure.  Depending  on  the  signature  of 4-space curvature our
solutions  admit  of indefinite expansion in the usual 3-space and
dimensional  reduction of the fifth dimension. Due to the presence
of  the  scalar  field  the  case  $p=-\rho$  does  not  yield  an
exponential  expansion  of  the  scale  factor,  which  strikingly
differs  from  our earlier investigations without scalar field.The
 \emph{effective} four dimensional values of entropy and matter are
 calculated  and  possible  consequences  of  entropy  and  matter
 generation  in  the 4D world as a result of dimensional reduction
 of      the      extra      space     are     also     discussed.
Encouraging  to  point out that aside from the well known big bang
singularity  our  inhomogeneous  cosmology  is  spatially  regular
everywhere.  Further  our  model  seems  to suggest an alternative
 mechanism    pointing    to    a    smooth    pass    over   from
a primordial, inhomogeneous cosmological phase to a 4D homogeneous
one.
\end{abstract}

   ~~~~~~~~KEYWORDS : cosmology; higher dimensions; scalar field

\bigskip
\section*{1. Introduction}

There has been of late a resurgence of interests in theories both
in general relativity and particle physics where the dimension of
space time is greater than the usual (3+1) we observe. These theories
include kaluza-klein , induced matter, super string,
supergravity and string. The interest in higher dimensional
spacetime also stems from its recent applications to brane
cosmology (see ref. 1 for a lucid exposition of these ideas)
\cite{wess}. In these (4+d) dimensional models the d-spacelike
dimensions are generally spontaneously compactified and the
symmetries of this space appear as gauge symmetries of the
effective 4D theory.

Homogeneous Kaluza-Klein extension of the FRW model has been
fairly adequately discussed in the literature \cite{rj}. Starting
from a topology of $R^{1}\times R^{3}\times S^{d}$ it is shown
that both the \emph{standard} and the extra space expand initially after
which the curvature effects become significant and the extra
compact space collapses to a singularity. It is conjectured that
some sort of quantum gravity effect produces a repulsive potential to
stabilize the compact space
at the planckian length and there after the visible universe
expands in the FRW way.

However, inhomogeneous cosmological models in higher dimensions
have not so far attracted the attention it deserves. But the
analysis  made by de Lapparent et al \cite{lap} of the \emph{cfa}
redshift  survey and also the observations of Saunders et al
\cite{sau} of Infrared Astronomy Satellite survey indicate that
the large scale structure of the universe does not show itself as
a smooth and homogeneous distribution of matter as was believed
earlier. At the same time the failure of the theoretical
considerations such as statistical fluctuations in the FRW models
to explain the large scale structure suggests that the
inhomogeneity factor in any physical cosmological model can no
longer be avoided. Motivated by these considerations some of us
have attempted to study inhomogeneous models and their
implications in a series of papers in 5D space time \cite{sc}.

On the other hand, it is known that the idea of inflation tries to
address the twin problems of horizon and flatness that plague the
standard big bang model. It is conjectured that the vacuum energy
driving the inflation is associated with a scalar field called
inflaton. The futile effort to identify the inflaton in the
particle sector and the fact that the gravitational force is the
only long range force governing the evolution of the universe
leads us to consider inflation as a pure gravitational effect.
This as well as the problem associated with the 'graceful exit'
criteria led people to seriously reconsider the Brans-Dicke
theory. The recent interest in B-D theory also stems from its
relation to low energy bosonic string theory. It is observed that
the two theories give identical predictions when the B-D coupling
constant $\omega = -1$. All these considerations have led to
several recent investigations into the generation of exact
solutions for Cosmology \cite{jdb}.

The present work is primarily motivated by our desire to
investigate the influence of B-D field in the frame work of
higher dimensional inhomogeneous space time assuming varied
equations of state. The paper is organized as follows: In section
2  the  field equations and their integrals are found where we
observe that as the 3D scale factor expands  the extra dimension
shrinks with time. This is desirable in the context of any higher
dimensional theory. In this section we also discuss the specific
cosmological evolution  assuming different equations of state. The
case $p=-\rho$ is strikingly different in the sense that here we
do not get exponential expansion of the scale factor as is
customary under identical situations. This may be due to the presence
of the scalar field. Here we have also studied
 the specifics of our model as regards the effective entropy generation, matter creation
 etc in the 4D world. The paper ends with a discussion in
section 3.

\section*{2. The Field Equations and Their Integrals}

The 5D line-element is taken as
\begin{eqnarray}
  ds^{2} &=&
  dt^{2}-R^{2}(dr^{2}+r^{2}d\theta^{2}+
  r^{2}sin^{2}\theta d\phi^{2}) - B^{2}dy^{2}
\end{eqnarray}
where $R=R(t)$ and $B=B(r,t)$ and $y$ is the fifth co-ordinate. So
here we assume that the physical 3-space is both flat and
homogeneous while inhomogeneity is being introduced through the
extra space. This, however, makes the total 5D space time an
inhomogeneous one. One should also note that since $B=B(r,t)$  we
are not dealing here with a product space, the shape of the extra space
is different at different points of the 4D world. We shall
subsequently see that the fact that inhomogeneity is being
introduced via the extra space has far reaching implications in
the cosmological evolution of our models.

We shall write down the field equations in Dicke's revised units
(Einstein frame)\cite{akr} as

\begin{equation}
G_{ij}=R_{ij} - \frac{1}{2} g_{ij}R= -( T_{ij}+\Lambda_{ij})
\end{equation}
\begin{equation}
\Lambda_{ij}= \frac{2\omega+3}{2\psi^{2}}(\psi_{i}\psi_{j} -
\frac{1}{2}g_{ij}\psi_{k}\psi^{k})
\end{equation}
where $\psi,_{i}\equiv \psi_{i}$

\begin{equation}
\Box (\log \psi)= \frac{T}{2\omega+3}
\end{equation}
Here $T_{ij}$ is the usual energy momentum tensor for the matter
field and $\Lambda_{ij}$ is that due to the scalar field in
revised units. In what follows we assume for simplicity that the scalar
field  depends on time only, i.e., $\psi = \psi(t) $. The only
non vanishing components of $\Lambda_{ij}$ are
\begin{equation}
-\Lambda_{0}^{0}=\Lambda^{1}_{1}=\Lambda^{2}_{2}=\Lambda^{3}_{3}= \Lambda^{4}_{4}
=-\frac{2\omega+3}{4}(\frac{\dot{\psi}}{\psi})^{2}=-\frac{k}{2}(\frac{\dot{\psi}}{\psi})^{2}
\end{equation}
where $k= \frac{2\omega+3}{2}$. In comoving co-ordinates the field
equations (2) for the metric(1) are given by
\begin{eqnarray}
G_0^1 &=& \frac{\dot{B'}}{B} - \frac{B'\dot{B}}{B^{2}}- \frac{\dot{R}B'}{RB} = 0   \\
G_1^1 &=& \frac{2}{R^{2}r}\frac{B'}{B} - (2\frac{\ddot{R}}{R} +
\frac{\dot{R^{2}}}{R^{2}} + \frac{\dot{B^{2}}}{B^{2}}+
2\frac{\dot{R}\dot{B}}{RB})= p +
\frac{k}{2}(\frac{\dot{\psi}}{\psi})^{2}   \\
G_{2}^{2}&=& G_{3}^{3}= \frac{1}{R^{2}}(\frac{B''}{B} +
\frac{1}{r}\frac{B'}{B}) - (2\frac{\ddot{R}}{R} +
\frac{\dot{R^{2}}}{R^{2}} + \frac{\dot{B^{2}}}{B^{2}} +
2\frac{\dot{R}\dot{B}}{RB})= p +
\frac{k}{2}(\frac{\dot{\psi}}{\psi})^{2}  \\
G_{4}^{4}&=& - 3(\frac{\ddot{R}}{R} +
\frac{\dot{R^{2}}}{R^{2}})=p_{5}
+\frac{k}{2}(\frac{\dot{\psi}}{\psi})^{2}  \\
G_{0}^{0}&=& \frac{1}{R^{2}}(\frac{B''}{B} +
\frac{2}{r}\frac{B'}{B} ) - 3(\frac{\dot{R^{2}}}{R^{2}} +
\frac{\dot{R}\dot{B}}{RB})= - \rho -
\frac{k}{2}(\frac{\dot{\psi}}{\psi})^{2}
\end{eqnarray}
and the wave equation is
\begin{equation}
\frac{\ddot{\psi}}{\psi}-\frac{\dot{\psi^{2}}}{\psi^{2}}+
3\frac{\dot{\psi}}{\psi}\frac{\dot{R}}{R}+\frac{\dot{\psi}}{\psi}\frac{\dot{B}}{B}=\frac{T}{2k}
\end{equation}
(where T is the trace of energy-momentum tensor).
 From equation(6)
\begin{equation}
B=\beta(r)R+\alpha(t) \end{equation} where $\beta(r)$ and
$\alpha(t)$ are integration function of r and t respectively, and
from equation (7) and (8)
\begin{equation}
B=-c\gamma(t)r^{2}+\alpha(t) \end{equation}  where c is a constant
of integration and $\gamma (t) $ is integration function of t. Now
comparing equation (12) and (13) we finally get
\begin{equation}
B=-cr^{2}R+\alpha(t) \end{equation}

\bigskip
\textbf{CASE I} ($p=p_{5}=0 $ : dust distribution )\\[1mm]

It is very difficult to solve the set of equations in a general
form. So in view of the fact that the power law expansion in BD
theory is now being seriously considered as a possible scenario to
solve the 'graceful exit' problem in inflationary cosmology we
assume $R \sim t^{a}$, where 'a' is a constant. Now eliminating
$\rho$ from equations(10) and (11) we get via equation
(14) the following two relations
\begin{equation}
3\frac{\dot{R^{2}}}{R^{2}}=k(\frac{\ddot{\psi}}{\psi}-\frac{3}{4}
\frac{\dot{\psi^{2}}}{\psi^{2}})
+4k\frac{\dot{R\dot{\psi}}}{R\psi}
\end{equation}
\begin{equation}
(3\frac{\dot{R}}{R}-2k\frac{\dot{\psi}}{\psi})\dot{\alpha}+
\{3\frac{\dot{R^{2}}}{R^{2}}-2k(\frac{\ddot{\psi}}{\psi}-
\frac{3}{4}\frac{\dot{\psi^{2}}}{\psi^{2}}+
3\frac{\dot{R}\dot{\psi}}{R\psi})\}\alpha+6\frac{c}{R}=0
\end{equation}
From equation (9) we also get
\begin{equation}
\psi=t^{\sqrt{\frac{6a(1-2a)}{k}}} \end{equation}  The equation (15)
further fixes the value of 'a' to be either $a=\frac{8k}{3+16k}$
or $a=\frac{1}{4}$. But in order to be consistent with other
equations the first value is to be discarded and also
$k=\frac{3}{4} $ (i.e., $\omega=-\frac{3}{4}$). For economy
 of space we skip all intermediate steps and solve equation
 (16) to find $\alpha $ as
\begin{equation}
\alpha(t)=bt^{\frac{1}{4}}+\frac{16c}{4\sqrt{3k}-3}t^{\frac{7}{4}}
\end{equation}
So all pieces put together finally yield
\begin{eqnarray}
  R&=& t^{\frac{1}{4}} \\
  B&=& (b-cr^{2})t^{\frac{1}{4}}-\frac{16c}{9}t^{\frac{7}{4}} \\
  \psi &=& \frac{1}{t} \\
  \rho &=& \frac{36c}{9(b-cr^{2)t^{\frac{1}{2}}}-16ct^{2}}
\end{eqnarray}

From the equation (22) it follows that $\rho > 0 $ implies that $ c > 0 $.
\begin{figure}
\resizebox{!}{4.5in}{\includegraphics{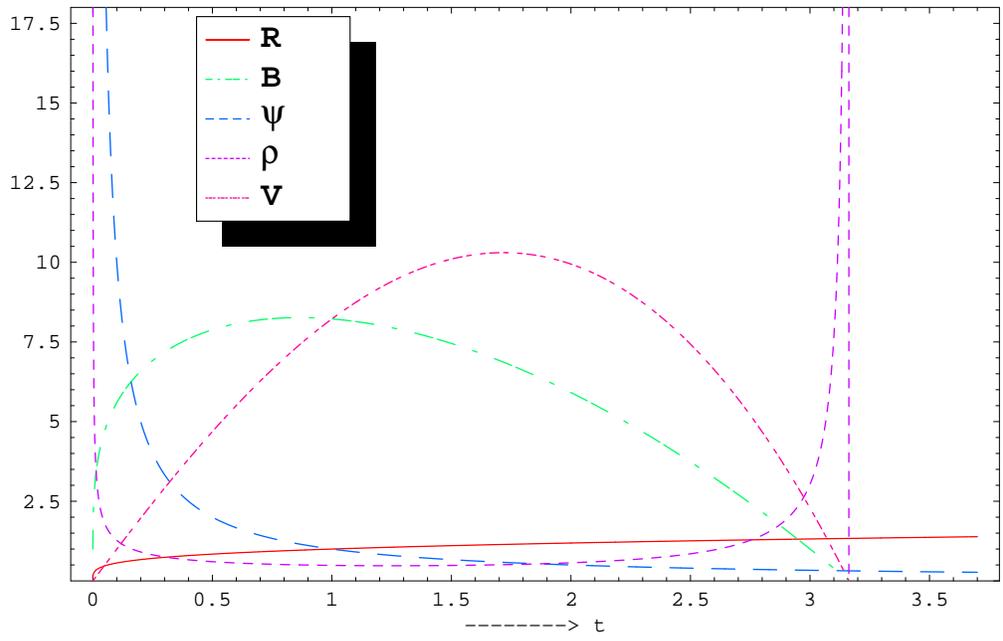}}
\caption{\small \it Plot of $R, B, \psi, \rho$ and $V$ vs time. The value of the parameter
$c$ has been taken to be 1, and $b = r^2 + 10$.}
\end{figure}
One can, at this stage, calculate the 4-space curvature of the
t-constant hyper surface for the line element (1). The relevant
expression is given by \cite{cb}
\begin{equation}
R_{i}^{i}=R^{*(4)}+\dot{\theta}+\theta^{2}-2\omega^{2}+\dot{u}^{i}_{;i}
\end{equation}
where $\theta$ is the expansion scalar and the last two terms
refer to vorticity and acceleration and $R^{*(4)}$ is the 4-space
curvature.

After some algebra we get
\begin{equation}
R^{*(4)}=\frac{12c}{(b-cr^{2})t^{\frac{1}{4}}-\frac{16c}{9}t^{\frac{7}{4}}}
\end{equation}

Thus we can fix up the arbitrary constant c as a measure of the
curvature of the 4-space. As is evident from the expression (24)
this curvature, unlike the homogeneous FRW model is a function of
the spatial co-ordinates also and it blows off at the initial
singularity t=0. Otherwise it is regular everywhere.
The shear scalar also comes out to be
\begin{equation}
\sigma^{2}=\sigma_{ij}\sigma^{ij}=\frac{3}{2}(\frac{\dot{B^{2}}}{B^{2}}
-\frac{\dot{R^{2}}}{R^{2}})=\frac{8c^{2}t^{\frac{3}{2}}}{3B^{2}}
\end{equation}
So with time the model increasingly anisotropises as the
denominator reduces to planckian length. This is only to be
expected in any sensible multidimensional model.

To check if our space time contains any geometric singularity
aside from the well known big bang one at t=0, we calculate the
Kretschmann scalar as $R_{ijkl}R^{ijkl}$. Explicit calculations
(also checked and verified with the help of a computer-aided
program) show that only the $R_{0441}$ survives such that the
scalar reduces to
\begin{equation}
R_{ijkl}R^{ijkl}=\frac{2}{R^{2}B^{2}}(B^{''})^{2}
\end{equation}
Nice thing  about it is that for our model the scalar is regular
everywhere including the point r=0, which may be identified with
the centre of the fluid distribution as in many other
inhomogeneous distributions.

\bigskip
\textbf{Dynamical Behaviour: }\\[1mm]

The behaviour is schematically shown in the adjoining figure (1).
The spatial volume is given as
\begin{equation}
V =R^{3}B=(b-cr^{2})t-\frac{16c}{9}t^{\frac{5}{2}}
\end{equation}
For positive curvature ($c>0$) the extra space collapses at a finite
time $t=t_{c}$. If the
extra dimension does collapse to zero in a finite time we can not
analytically continue time beyond that in the usual case. A
possible way out of the difficulty is to find out some sort of
stabilizing mechanism which might fix up the extra scale at a
planckian length.

Moreover, if we consider the time evolution of a shell
characterized by a r-const. hyper surface it follows that for $c
>0$, the 4-volume $(R^{3}B)$ collapses. This also follows from
equation (24) as $R^{*(4)}>0$ for $c >0$. Interestingly both the
spatial volume and the extra dimension collapse at the same
instant (fig 1).

One difference from the analogous homogeneous case is too obvious.
While in the homogeneous case the whole distribution collapses at
a single instant, here the collapse time depends on the radial
co-ordinate. So the big crunch is not simultaneous for all the
shells.

\bigskip
\textbf{Entropy Generation: }\\[1mm]

Before concluding the section let us attempt to address a
particular problem of current interest - the very high value of
entropy per baryon observable at present. If one assumes that the
4D formulation of the laws of thermodynamics can be extended to 5D
also and the interactions at the early universe led to
thermodynamical equilibrium, the total entropy of the universe in
an adiabatic change can not but conserve. Now $S_{4}$,
the effective four dimensional entropy that we observe at present is given by
\begin{equation}
T\frac{dS_{4}}{dt}=\frac{d}{dt}(\rho R^{3})+p\frac{d}{dt}(R^{3})
\end{equation}
Again divergence of equation (2) gives
\begin{equation}
\dot{\rho}+(\rho+p)\frac{\dot{3R}}{R}+(\rho+p_{5})
\frac{\dot{B}}{B}+\frac{1}{2}(\rho-3p-p_{5})\dot{\psi}=0
\end{equation}
such that
\begin{equation}
T\dot{S_{4}}=-(\rho+p_{5})\frac{\dot{B}}{B}-\frac{1}{2}(\rho-3p-p_{5})\dot{\psi}
\end{equation}
In our model both $\dot{B}$ and $\dot{\psi}$ are negative giving
$\dot{S_{4}}>0$ for physically reasonable matter field. So the
observable entropy increases as  the extra dimension shrinks.And
the effect is augmented due to the presence of the scalar field.
The above situation is discussed for homogeneous distribution
without scalar field by Alvarez and Gavela \cite{ag} in an earlier
work in the context of a homogeneous model.

\bigskip
\textbf{Matter Leakage:}\\[1mm]

One may look at the conservation relation (29) from another stand
point also. We know that to avoid big bang singularity Gold and
Bondi \cite{gb} hypothesized continuous creation of matter without
offering any dynamical mechanism for it. Later Hoyle and Narlikar
\cite{hn} invoked an extraneous 'creation field' in their C-field
theory for matter creation. In both the cases the conservation
principle is clearly given up. It may be tempting to suggest an
alternative scenario where conservation principle strictly holds
in the 5D world.
 For our dust case and in absence of scalar field the equation (29)
  leads to
\begin{equation}
\rho R^{3}B= Constant=M(0)
\end{equation}
where M(0) may be identified with the total mass in the 5D world
which strictly conserves. But for a 4D observer the effective 4D
matter is given by $\rho R^{3}=M_{4}$, such that
\begin{equation}
M_{4}(t)=M_{0}B^{-1} \end{equation} Since for a realistic model
the extra dimension shrinks the above equation tells us that
although the overall (4+1)D matter remains conserved there will be
matter leakage from the internal space onto the effective 4D
world. So it offers a natural mechanism of matter creation in 4D
space time without the assumption of an extraneous field
\cite{dt}as also without violating any conservation principle
in physics.

\bigskip
\textbf{Case II :~  $p\neq p_{5}\neq 0$}\\[1mm]

While working with matter field in higher dimensional theories
three possibilities present themselves : (i) a vacuum state, (ii)
a low temperature state, i.e. , $T<\frac{1}{B}$ where evidently
$\frac{1}{B}$ gives the curvature scale of $S^{1}$, and (iii) a
high temperature state, i.e. , $T>\frac{1}{B}$. In this case we
take the third possibility where the energy of the particles is
higher than the excitation energy of the internal space such that
the higher excitations in the internal space can no longer be
neglected. Therefore, the higher dimensional stress $p_{5}\neq 0$
and the isometry of our metric dictates that in the comoving
co-ordinates $T^{1}_{1}=T^{2}_{2}=T^{3}_{3}= -p$. In this case the
wave equation is
\begin{equation}
\frac{\ddot{\psi}}{\psi}-\frac{\dot{\psi^{2}}}{\psi^{2}}+
(3\frac{\dot{R}}{R} +
\frac{\dot{B}}{B})\frac{\dot{\psi}}{\psi}=\frac{\rho-3p-p_{5}}{2k}
\end{equation}
\begin{figure}
\resizebox{!}{4.5in}{\includegraphics{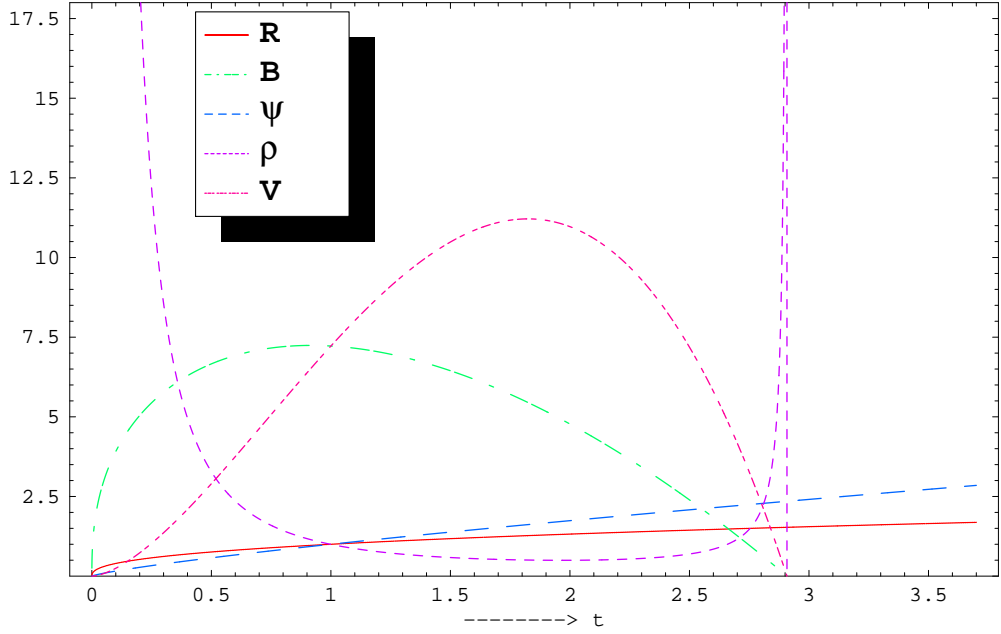}}
\caption{\small \it Plot of $R, B, \psi, \rho$ and $V$ vs time. The value of the parameter
$c$ has been taken to be 1, $k = 3/4$ and $b = r^2 + 10$.}
\end{figure}
For simplicity we assume
$\frac{\dot{\psi}}{\psi}=2\frac{\dot{R}}{R}$ and for economy of
space we skip all intermediate mathematical steps to finally  get
\begin{eqnarray}
  R &=& t^{\frac{2}{5}} \\
  B &=& (b-cr^{2})t^{\frac{2}{5}}-\frac{75 c}{33-8k}t^{\frac{8}{5}} \\
  \psi &=& t^{\frac{4}{5}} \\
  p &=& \frac{(\frac{6-8k}{25})(b-cr^{2})t^{-\frac{8}{5}}+
  \frac{(12+56k)}{33-8k}ct^{-\frac{2}{5}}}{(b-cr^{2})t^{\frac{2}{5}}-
   \frac{75c}{(33-8k)}t^{\frac{8}{5}}}\\
  \rho &=& \frac{(\frac{24-8k}{25})(b-cr^{2})t^{-\frac{8}{5}}+
  \frac{(18- 24k)}{33-8k}ct^{-\frac{2}{5}}}{(b-cr^{2})t^{\frac{2}{5}}-
  \ \frac{75 c}{(33-8k)}t^{\frac{8}{5}}}\\\\
  p_{5}&=& \frac{6-8k}{25}t^{-2}
\end{eqnarray}
the only restriction being $(2\omega + 3)<\frac{3}{2}$.

The temporal behaviour in this model is shown in figure (2). Here
the 3D space expands indefinitely and the extra dimension is
amenable to reduction. Further if the 4D curvature $c=0$ we get
$R=B$ and the model becomes isotropic and homogeneous and
interestingly $p=p_{5}$. Further to mention that when
$k=\frac{3}{4}$ the five dimensional pressure vanishes.

\bigskip
\textbf{Case III :~ $p=-\rho$,  $p_{5}=0$} \\[1mm]

From equation (7) and  (10),
\begin{eqnarray}
  \frac{\ddot{R}}{R}-\frac{\dot{R^{2}}}{R^{2}} &=& -\frac{k}{3}
  (\frac{\dot{\psi}}{\psi})^{2} \\
  \ddot{\alpha}-\frac{\dot{R}}{R}\dot{\alpha}+ \{2(\frac{\ddot{R}}{R}
  -\frac{\dot{R^{2}}}{R^{2}})+k(\frac{\dot{\psi}}{\psi})^{2}\}\alpha&=& \frac{2c}{R}
\end{eqnarray}
Again as $p_{5}=0$, we get from equations (41) and (9)
\begin{eqnarray}
  R &=& t^{\frac{1}{4}} \\
  \psi &=& t^{\sqrt{\frac{3}{4k}}}
\end{eqnarray}
while equation (42) yields
\begin{equation}
\alpha= bt^{\frac{1}{4}}+\frac{16c}{9}t^{\frac{7}{4}}
\end{equation}
Here also we have a restriction on k as $k= 48$, i.e. , $\omega=
\frac{93}{2}$ and
\begin{eqnarray}
B&=&(b-cr^{2})t^{\frac{1}{4}}+\frac{16 c}{9}t^{\frac{7}{4}}\\
p&=&-\rho=-\frac{72 c}{9(b-cr^{2})t^{\frac{1}{2}}+16ct^{2}}
\end{eqnarray}

The equation (47) further dictates that as $\rho$ and B are non negative
the constant c should be greater that zero.Hence dimensional
reduction in this model is clearly ruled out.There are two
striking features in this case. Firstly the $'\rho'$ and $'p'$ are
not separately constant as in 4D case. This, however, follows from
the divergence relation (29). Secondly there is no exponential
expansion of the 3D scale factor. In our earlier work \cite{sc} we
have got exponential expansion of the 3D scale factor in 5D model
without any scalar  field. So the presence of the scalar field
makes the difference.

\bigskip
\textbf{3. Discussion :}\\[1mm]

Before concluding let us sum up the basic features of our model.
Here we consider an inhomogeneous 5D cosmology where the
inhomogeneity is tacitly introduced through the extra fifth
dimension. Encouraging to point out that apart from the well known
big bang singularity we do not encounter any spatial singularity
including the centre of symmetry (r=0) as evidenced from the
regularity of the kreschmann scalar. Secondly the extra dimension
exhibits dimensional reduction, a property which is always sought
for while working with multidimensional cosmology.

Moreover,the recent COBE DMR experiments \cite{smoot} suggest a scale of
temperature fluctuation, which may be related to density
fluctuation, thus providing unambiguous evidence for the existence
of inhomogeneity in the early universe. Since the visible world
around us is manifestly homogeneous and isotropic, various models
have been proposed to account for a possible pass over from the
primordial inhomogeneous scale to the current homogeneous one.
Explicit computations \cite{barrow} however, showed that the so
called \emph{'chaotic'} models \cite{misner} can not explain these
observed cosmological properties. The scenario presented in this
work seem to provide an alternative resolution to this problem as
follows : It is believed that the extra dimension once contracting
will ultimately stabilize at a very small length and after that
the extra dimension will lose its dynamical character. But an
unphysical nature of this model is that it is not apparent from our
analysis how that stabilization works. However, if the number of
extra dimensions is at least two it may generate a repulsive
potential to halt the contraction at a certain stage \cite{gk}.

Whatever be the stabilizing mechanism if one believes that it does
occur then for our model it has far reaching consequences. Because as
the fifth dimension becomes constant then our model becomes
exactly FRW type. So not only do we enter a 4D world it also
envisages a smooth transition from a 5D inhomogeneous cosmology to
4D homogeneous one. And this homogenization takes place without
forcing us to choose very special initial conditions as is
customary in other standard 4D models.

To end a final remark may be in order.In some of the cases discussed
 above the B-D coupling constant $\omega$ comes out to be negative.
 This is not as undesirable aspect as was thought to be earlier.
 Because as pointed out in the introduction  the low energy sector
  of the string theory contains an action which may be identified
  with the B-D action subject to the restriction that $\omega = -1$.
  Further the much discussed quintessence phenomenon in the frame-
  work of scalar tensor theory suggests that for the accelerated
  expansion of the Universe the $\omega$ should be negative even when an
  additional potential field is introduced in the action.

\bigskip
\textbf{Acknowledgment : }

 S. C. wishes to thank TWAS, Trieste for
travel support and ITP (Beijing) for local hospitality where the
work was initiated. The financial support of UGC is also
acknowledged.


\begin{thebibliography}{22}

\bibitem{wess}Wesson P. S.(1999) Space-Time-Matter, World
Scientific, Singapore

\bibitem{rj}Randjbar-daemi S., Salam, A. and Strathdee, J.
(1984) Phys. Lett.\textbf{135B},388; Chatterjee S. and Bhui B.
(1990) Mon. Not. R. Ast. Soc. \textbf{108},252

\bibitem{lap}de Lapparent G., Geller M. and Huchra J. P.
(1986)Astrophys. J.\textbf{302}.L1

\bibitem{sau}Saunders W. et al (1991) Nature \textbf{349},32

\bibitem{sc}Banerjee A., Panigrahi D.,Chatterjee S. (1994) Class. Quntum Grav.\textbf{11},1405;
Chatterjee S., Beesham A. Bhui B. and Ghosh T. (1998) Phys. Rev
\textbf{D57}, 6544

\bibitem{jdb}Barrow J. D. (1993) Phys. Rev. \textbf{D47}, 5329; Damour T. and
Nordvedt K. (1993) Phys. Rev. \textbf{D48}, 3436

\bibitem{akr}Raychowdhuri A. K., Theoretical Cosmology, Oxford
University press, 1979
\bibitem{cb}Chatterjee S. and Banerjee A.
 (1993) Class. Quntum Grav. \textbf{10}, L1
\bibitem{ag}Alvarez E. and Gavela M. (1983) Phys. Rev. Lett. \textbf{51},
931
\bibitem{gb}Bondi H., and Gold T. (1948) Mon. Not. R. Ast. Soc \textbf{108}, 252
\bibitem{hn}Hoyle F. and Narlikar J. V. (1966)Proc. Roy. Soc. \textbf{A290}, 162
\bibitem{dt}Dereli I. and Tucker R. W. (1983) Phys. Lett. \textbf{125B},
133

\bibitem{smoot}Smoot G. F. et al (1992) Astrophys. J. \textbf{396}, L1

\bibitem{barrow}Barrow J. D. (1982) Phys. Rep. \textbf{85},1


\bibitem{misner}Misner C. W. (1968) Astrophys. J. \textbf{151}, 453

\bibitem{gk}Guendelman E. I. and Kaganovich A. B. (1993) Int. J.
Mod. Phys. \textbf{D2}, 221


\end{thebibliography}
\end{document}